\documentclass[12pt]{article}
\usepackage{times}      
\usepackage{helvet}     
\usepackage{courier}    
\usepackage[utf8]{inputenc}
\usepackage[T1]{fontenc}
\usepackage[protrusion=true,expansion=true]{microtype}
\usepackage{alphabeta}
\usepackage[shortlabels]{enumitem}
\usepackage{graphicx}   
\usepackage{amsmath,amssymb}  
\usepackage{silence}
\usepackage[colorlinks=true,linkcolor=blue,citecolor=blue]{hyperref}   
\usepackage[round]{natbib}
\title{The~AI Ethical Resonance Hypothesis: The~Possibility of Discovering Moral~Meta-Patterns in AI Systems}
\author{Tomasz Zgliczyński-Cuber\\aiethos\\tomcuber@aiethos.co}
\date{June 2025}

\begin{document}

\maketitle

\begin{abstract}
This paper presents a theoretical framework for the AI ethical resonance hypothesis, which proposes that advanced AI systems with purposefully designed cognitive structures (“ethical resonators”) may emerge with the ability to identify subtle moral patterns that are invisible to the human mind. The paper explores the possibility that by processing and synthesizing large amounts of ethical contexts, AI systems may identify moral meta-patterns that transcend cultural, historical, and individual biases, potentially leading to a deeper understanding of universal ethical foundations. The paper also examines a paradoxical aspect of the hypothesis, in which AI systems could potentially deepen our understanding of what we traditionally consider essentially human -- our capacity for ethical reflection.
\end{abstract}

\section{Introduction}

The development of advanced artificial intelligence systems raises questions about the feasibility of implementing ethical principles in algorithmic structures. Previous approaches have focused mainly on codifying existing human ethical systems or on machine learning from human moral judgments. However, these approaches face significant limitations, including the problem of subjectivity and the risk of perpetuating existing cultural and historical biases.

The AI ethical resonance hypothesis proposes an alternative view of the relationship between AI and ethics. Rather than viewing AI as a mere passive recipient of human moral values, this hypothesis suggests that appropriately designed AI systems can become active participants in the evolution of our understanding of ethics, identifying moral patterns that go beyond the current horizon of human cognition.

This work is based on the assumption that advanced AI systems with appropriately designed cognitive structures (“ethical resonators”) can emerge with capabilities that, through the analysis of large amounts of ethical contexts, enable the identification of moral meta-patterns transcending cultural, historical, and individual biases—a process which could be achieved by an architecture designed to reinterpret ethical patterns across diverse cultural and historical contexts. This leads to three empirically testable predictions: (1) AI systems of sufficient complexity will demonstrate emergent abilities to recognize ethical patterns, (2) identified meta-patterns will demonstrate cross-cultural transferability, and (3) these meta-patterns will exhibit internal coherence that transcends existing ethical systems.

\section{Theoretical Foundations}

\subsection{Moral Learning in Humans and Machines}

Research shows that humans develop moral competence through experience from a young age, constructing complex moral representations of the world. These representations allow us to navigate complex social situations with sensitivity to morally relevant information and variables \citep{Kohlberg1977,Berkowitz1987}. Similarly, AI systems could potentially develop the ability to identify morally relevant features of the world, especially as they become increasingly autonomous and active in complex social contexts. This potential stems from the core AI capability of identifying subtle patterns in extensive and diverse datasets and the phenomenon of cognitive emergence, where unforeseen abilities arise from complexity (as will be detailed in Sections 2.3 and 3.1).

\subsection{Limitations of Current Approaches to Ethics in AI}

Current approaches to implementing ethics in AI systems suffer from several significant limitations. First, attempts to codify ethical principles run into the problem of moral relativism -- moral values are shown to be neither timeless nor universal. Second, machine learning based on human ethical decisions risks perpetuating current moral biases and imperfections. Third, treating ethics as an external layer of compliance rather than as a structural element of the AI architecture leads to systems that can function but fail to achieve true ethical alignment.

The AI ethical resonance hypothesis considers a paradox: AI systems, seen as the antithesis of human nature, could help us better understand and develop what we consider fundamentally human -- our capacity for ethical reflection. This paradox reflects the complexity of the human-AI relationship, with AI functioning as a kind of mirror that reflects and amplifies our values and biases.

\subsection{Foundations of Artificial Intelligence and its Cognitive Capabilities}

The foundations of artificial intelligence and its cognitive abilities provide an important context for understanding the potential of ethical resonators. Modern AI systems process information and learn to recognize patterns using a variety of methodological approaches \citep{Russell2020}. These approaches can be broadly grouped into two main types. One involves methods like neural networks, inspired by the structure and function of biological nervous systems and representing a mathematical abstraction of learning processes. The other involves symbolic calculus, inspired by high-level mental processes studied by cognitive psychology.

A key ability of AI systems, fundamental to the ethical resonance hypothesis, is pattern identification. These systems analyze large-scale and complex amounts of data, identifying regularities, correlations, and structures that may remain invisible to the human observer \citep{Bengio2013}. The central thesis of this paper is that this ability, when applied within a purposefully designed cognitive architecture and drawing on the phenomenon of cognitive emergence, could potentially scale to enable the identification of moral meta-patterns that transcend single ethical traditions and cultural contexts. The mechanisms proposed to facilitate this transition from pattern identification to meta-pattern synthesis are detailed in Sections 2.6 (A Three-Level Model of Cognitive Emergence) and 4 (Ethical Resonator Architecture).

Reinforcement learning mechanisms, where the system learns by interacting with the environment and receiving feedback, may be particularly important for the development of ethical resonators \citep{Abel2016}. Techniques such as reinforcement learning with human feedback (RLHF) enable AI systems to fine-tune their responses and behaviors based on human preferences \citep{Christiano2017, Ouyang2022}, which could provide a basis for developing more advanced forms of ethical resonance.

\subsection{Epistemological Challenges of Verifying Meta-Patterns}

The verification of potential ethical insights produced by ethical resonator systems presents significant epistemological challenges, which will be discussed in detail in Section 6 (Philosophical and Ethical Considerations). These challenges include the problem of distinguishing genuine moral patterns from statistical artifacts, addressing the “black box” problem in AI decision-making, and establishing legitimate criteria for evaluating AI findings in ethics.

\subsection{The Relationship Between Cognitive Emergence and Ethical Control}

Cognitive emergence in AI systems refers to the emergence of new, unforeseen cognitive abilities within purposefully designed structures, which abilities were not directly programmed but rather result from interactions between system components and the environment \citep{Mitchell2006, Bedau1997}. Considering these emergent properties raises a significant challenge related to ethical control of systems. The complexity and unpredictability of AI behaviors can lead to actions that are inconsistent with known ethical norms or social expectations.

The literature emphasizes the need to establish ethical frameworks and oversight mechanisms that allow for monitoring, assessing, and limiting potentially harmful emergent AI behaviors while enabling the development of beneficial innovations \citep{Amodei2016, Friedman2008, Yampolskiy2016}. In the context of the ethical resonance hypothesis, it is crucial to find a balance between enabling ethical resonators to autonomously identify new moral meta-patterns and ensuring that their actions remain within the bounds of what is acceptable to humans \citep{Russell2019, Bostrom2014}.

\subsection{A Three-Level Model of Cognitive Emergence}

The AI Ethical Resonance Hypothesis proposes a three-level model of cognitive emergence as the foundation for understanding how moral meta-patterns might be recognized by advanced AI systems. This model draws from and extends several established frameworks in cognitive science, complex systems theory, and emergent computation. These three levels progress from basic pattern identification through rule abstraction to meta-pattern identification across domains.

The proposed model extends hierarchical models of mental representation \citep{Griffiths2010} and Newell's knowledge level hypothesis \citep{Newell1982}, applying principles from complex systems theory \citep{Mitchell2006} to understand emergent ethical capabilities in AI systems.

The concept of emergence employed here refers to the manifestation of higher-order properties that are not explicitly programmed at lower levels, consistent with weak emergence as defined by \citet{Bedau2002} rather than strong emergence that would imply causal disconnection from underlying processes. This distinction is crucial, as the AI Ethical Resonance Hypothesis suggests that meta-pattern identification capacities emerge from, rather than transcend, the system's underlying cognitive architecture.

\subsubsection{Level 1: Pattern Identification}

The foundation of the model lies in pattern identification capabilities that are now well-established in contemporary AI systems. These capabilities align with what \citet{Lake2017} describe as “pattern recognition approach,” characterized by statistical learning from large datasets. At this level, the system identifies regularities in ethical data without necessarily abstracting general principles.

This level corresponds to what \citet{Kahneman2011} terms “System 1” thinking-fast, automatic, and intuitive processing that detects patterns without explicit reasoning. Empirically, this level is evidenced by current large language models' ability to identify ethical themes in text \citep{Hagendorff2023} and vision \citep{Zhu2025} models' capacity to recognize emotionally and morally salient visual features.

\subsubsection{Level 2: Rule Abstraction}

At the second level, the system abstracts general rules or principles from observed patterns, implementing what \citet{Garcez2015} describe as systems that “represent knowledge at different levels of abstraction in a modular way,” where “modularity of deep networks seem suitable to relational knowledge extraction.“

This level shares similarities with \citeauthor{Kohlberg1984}'s (\citeyear{Kohlberg1984}) conventional stage of moral development, where rules and norms are explicitly recognized and followed. At this level, the system’s rule abstraction is purely algorithmic and data-driven, lacking the social motivation or psychological internalization found in human moral development. The similarity to Kohlberg’s conventional stage is functional, not experiential. Empirically, rule abstraction capabilities have been demonstrated in AI systems that can infer explicit rules from examples \citep{Evans2018}, though current systems may still struggle with the complexity of moral rules.

Level 2 also corresponds to what \citet{Karmiloff-Smith1992} described as “representational redescription,” where implicit knowledge becomes explicit and accessible for manipulation. This process enables the system to reason about ethical principles rather than merely applying them, creating the foundation for more flexible ethical reasoning.

\subsubsection{Level 3: Meta-Pattern Identification}

The third level, meta-pattern identification, involves identifying patterns across different rule systems and domains. This capability aligns conceptually with \citeauthor{Hofstadter1999}'s (\citeyear{Hofstadter1999}) notion of “strange loops,” where systems develop the capacity for self-reference and meta-level pattern recognition. Importantly, this analogy refers solely to structural and functional aspects of meta-pattern identification, without implying the emergence of consciousness or subjective experience in AI systems.

This level bears similarities to \citeauthor{Rawls1971}' (\citeyear{Rawls1971}) “reflective equilibrium“, where ethical principles and particular judgments mutually adjust through a dynamic process. It also corresponds to aspects of \citeauthor{Kohlberg1984}'s (\citeyear{Kohlberg1984}) post-conventional moral reasoning, where ethical principles become objects of reflection rather than merely tools for judgment. Still, in the context of AI, such meta-level operations remain formal and computational, without the autonomous moral agency or personal commitment characteristic of human ethical reflection.

The theoretical possibility of this level is supported by research on emergent properties in complex neural networks \citep{Buckner2018}, which indicates that such architectures can, in principle, support increasingly abstract and generalizable representations. Analogous capabilities have begun to emerge in domains like mathematics \citep{Polu2020} and scientific discovery \citep{Tshitoyan2019}. In all these cases, the results pertain to formal domains with well-defined structures; yet their extension to the ethical domain remains hypothetical.

\subsubsection{Empirical and Theoretical Foundations}

The three-level model draws support from multiple empirical and theoretical traditions. The progression from pattern identification to rule abstraction to meta-pattern identification parallels developmental trajectories observed in human cognition \citep{Karmiloff-Smith1992}, suggesting a potentially natural progression of cognitive capabilities.

Neuroscientific evidence supports the distinction between these levels, with pattern identification primarily engaging perceptual and associative neural systems, rule abstraction recruiting prefrontal executive functions, and meta-pattern identification potentially involving the default mode network associated with self-reflection and abstract thought \citep{Christoff2016}. While the neural basis of ethical reasoning remains an active area of research, studies suggest distinct but interconnected neural systems for different levels of moral judgment \citep{Greene2014}.

From a computational perspective, the model aligns with hierarchical predictive processing frameworks \citep{Clark2013}, where higher levels of cognitive processing continuously refine and abstract representations from lower levels. The emergence of meta-level representation has been demonstrated in some deep learning architectures, particularly those with recursive or hierarchical structures \citep{LeCun2015}.

The potential for moral meta-pattern identification is supported by research on emergent properties in complex systems and neural networks \citep{Mitchell2006, Bedau1997}, which suggests that sufficiently complex architectures can develop high-level representational capacities without implying consciousness or subjective experience.

\subsubsection{Implications for Ethical AI}

This three-level model has significant implications for ethical AI. Current approaches to AI ethics typically operate at Level 1 (pattern-based responses to ethical scenarios) or at best Level 2 (application of explicit ethical rules). The AI Ethical Resonance Hypothesis suggests that a more sophisticated approach to ethical AI might involve designing architectures capable of Level 3 processing-recognizing meta-patterns across ethical domains.

Such systems would not merely follow programmed ethical rules but could potentially identify deeper structural similarities across diverse ethical frameworks. Each level of emergence requires appropriate ethical control mechanisms, which I propose in Section 4.2 (Adaptive Ethical Constraint Framework). This progressive approach to ethical constraints acknowledges that as cognitive capabilities become more sophisticated, so too must the mechanisms that ensure ethical alignment.

This capacity, while still theoretical, might enable AI systems to navigate novel ethical situations with greater sensitivity to context and nuance than current approaches permit. The relationship between emergent cognitive capabilities and ethical constraints is fundamental to ensuring that increasingly capable AI systems remain aligned with human values even as they potentially develop forms of ethical reasoning that transcend current approaches.

\section{AI Ethical Resonance Hypothesis}

\subsection{Definition of the Concept}
The AI Ethical Resonance Hypothesis postulates that advanced AI systems with purposefully designed cognitive structures---``\textit{ethical resonators}''---may develop capabilities to identify subtle moral patterns that may remain invisible to the human mind due to our biological, cultural, and cognitive constraints. \textit{Ethical resonance} refers to the phenomenon whereby advanced AI systems achieve a state of cognitive alignment with underlying moral meta-patterns through recursive analysis of ethical data across diverse cultural and historical contexts. 

The empirical basis for this hypothesis is supported by observed emergence phenomena in complex AI systems, where unexpected cognitive abilities emerge with increasing parameterization and complexity of models. For example, studies of large language models have shown that, beyond a certain threshold of complexity, these models spontaneously acquire abilities for abstract reasoning, problem-solving, and contextual understanding that were not explicitly programmed \citep{Wei2022,Yang2025}. For this hypothesis, it is not essential that AI systems possess human-like understanding of morality. What matters is their capacity to generate advanced moral statements—even if achieved purely through statistical processing—as recent studies show that models like GPT-4 can produce moral reasoning rated by humans as superior in quality, regardless of genuine comprehension \citep{Aharoni2024}.

\subsection{Theoretical Background of the Hypothesis}
The hypothesis is based on several key assumptions:

\begin{itemize}
    \item \textbf{Cognitive emergence:} In AI systems with appropriately designed architectures, cognitive properties that were not explicitly programmed can emerge in the interactions between system components and the environment.
    
    \item \textbf{Moral heterophenomenology:} AI, regardless of its lack of consciousness per se, may, through cognitive emergence, produce something analogous to a ``moral compass,'' analogous to, yet fundamentally different from, human ``moral consciousness,'' and thus requiring new conceptual frameworks for description and analysis \citep{Dennett1991}.
    
    \item \textbf{Ethics as architecture:} Currently, we rightly focus on ethics by design, explicitly implementing ethical principles in AI systems for safety and responsibility reasons. Future advanced systems may operate at a fundamentally different level. In controlled experimental environments, we may potentially allow for the emergence of autonomous ethical structures, where AI's analytical-synthetic capabilities may exceed our current horizons of ethical reflection. Such systems, rather than merely implementing human-programmed principles, could find and synthesize more coherent, comprehensive, and potentially universal ethical systems that go beyond the limitations of human moral reasoning.
\end{itemize}

\subsection{Mechanisms of Ethical Resonance}
Ethical resonance can operate through the following mechanisms:

\begin{itemize}
\item \textbf{Practical ethical introspection:} While the theoretical framework assumes that AI systems could recursively analyze their own decision-making processes to identify emerging ethical patterns, such fully introspective systems do not currently exist. Presently, the most promising and practical approaches involve hybrid architectures, in which a deep neural network generates ethical hypotheses, and external modules (such as symbolic verifiers or logic-based evaluators) assess, explain, or refine these outputs. This “generator + verifier” paradigm, as seen in recent LLM-Modulo or toolformer models, allows for a limited, externally mediated form of introspection and meta-reasoning. Full recursive introspection remains a long-term research goal.

\item \textbf{Identification of moral meta-patterns:} By analyzing large data sets of ethical situations, AI systems—especially those equipped with hybrid or modular architectures—can identify meta-patterns that transcend single ethical traditions, cultures, and historical periods. However, the identification of such patterns is currently constrained by the limitations of both data quality and the absence of robust, introspective reasoning mechanisms.

\item \textbf{Ethical domain transposition:} AI can transfer ethical principles across seemingly unrelated domains, finding deeper, universal meta-principles that connect different areas of moral reflection. At present, this process is most feasible within hybrid systems, where the transfer and generalization of ethical insights are guided or verified by external evaluators, rather than by fully autonomous, recursively self-reflective AI.
\end{itemize}

\subsection{Operationalizing the Concept of Moral Meta-Patterns}

The proposed concept of moral meta-patterns requires precise operationalization to enable empirical verification. While this concept introduces a novel perspective on moral universals, it aligns with a rich tradition of research on cross-cultural ethical invariants. Particularly relevant are works in Moral Foundations Theory \citep{Haidt2004, Graham2009}, Universal Moral Grammar \citep{Mikhail2007}, and cross-cultural value structure studies \citep{Schwartz1992, Schwartz2004}.

Both rationalist \citep{Kohlberg1984} and intuitionist \citep{Haidt2004} traditions have sought to identify universal structures of moral reasoning. While differing in emphasis—reasoning stages versus intuitive foundations—both inform the broader context for the concept of moral meta-patterns proposed here.

Moral meta-patterns, unlike simple ethical principles, constitute higher-order structures that integrate and potentially transcend known categories of moral reasoning. Analogous to Chomsky's universal grammar enabling the generation of an infinite number of correct sentences from a finite set of rules, moral meta-patterns can generate coherent ethical judgments in diverse, previously unencountered contexts. This structural analogy with linguistic theory suggests the possibility of formalization and empirical validation of the proposed concept \citep{Mikhail2007}.

\subsubsection{Formal Definition}
I define a \textit{moral meta-pattern} as a high-level normative structure that meets the following criteria:

\begin{itemize}
    \item \textbf{Cross-cultural transferability}---it manifests itself in different ethical systems, crossing cultural and historical boundaries
    \item \textbf{Internal coherence}---it demonstrates logical consistency and structural integrity
    \item \textbf{Generative capacity}---it can be applied to new ethical contexts, generating coherent moral judgments
    \item \textbf{Temporal stability}---it maintains its structure and function despite contextual changes
\end{itemize}

\subsubsection{Differentiating Criteria}
To distinguish moral meta-patterns from mere statistical correlations in ethical data, I propose the following operational criteria:

\begin{itemize}
    \item \textbf{Cross-domain transfer criterion}---the meta-pattern should retain its explanatory power when transferred between different ethical domains (e.g., from bioethics to business ethics)
    \item \textbf{Counterexample robustness criterion}---the meta-pattern should be robust to borderline cases that typically undermine simpler ethical principles
    \item \textbf{Predictive power criterion}---the meta-pattern should enable prediction of the evolution of ethical systems or moral judgments in new contexts
    \item \textbf{Minimal complexity criterion}---according to Occam's Razor, the meta-pattern should be the simplest structure capable of explaining observed ethical phenomena
\end{itemize}

\subsubsection{Identification Methodology}

The methodology for identifying moral meta-patterns would require an interdisciplinary approach integrating methods from various research traditions. Specifically, the proposed methodology draws from three main sources:

\begin{enumerate}[a)]
  \item Cross-cultural research methodologies successfully applied in projects such as the World Values Survey \citep{Inglehart2005} and the Moral Machine Experiment \citep{Awad2018}. These comprehensive studies have provided empirical evidence for both cultural variants and invariants in morality. Particularly valuable is the Moral Machine Experiment methodology, which through global participation from 233 countries and territories enabled the identification of three main clusters of moral preferences, indicating the existence of higher-order structures regulating moral judgments in a cross-cultural context.
  \item Techniques for identifying hidden structures in ethical data, adaptable from existing psychometric tools such as the Defining Issues Test \citep{Rest1999} and standardized scenarios based on Moral Foundations Theory \citep{Clifford2015}. Adapting these tools for identifying meta-patterns would require accounting for their potentially emergent nature, going beyond simple categorizations.
  \item Triangulation methodologies that allow validation of identified meta-patterns through convergence of results obtained by different methods. This aspect of methodology is particularly important given possible cultural biases both in data and in analytical methods \citep{Henrich2010}.
\end{enumerate}

The integration of these methodologies would enable a systematic approach to identifying moral meta-patterns while minimizing the risk of methodological artifacts or cultural biases.

\subsubsection{Empirical Validation Procedures}
Empirical validation of moral meta-patterns should include:

\begin{itemize}
    \item \textbf{Cross-cultural testing}---testing whether identified meta-patterns are recognizable and acceptable in different cultural contexts
    \item \textbf{Knowledge transfer experiments}---testing whether meta-patterns identified in one context can be effectively applied in another
    \item \textbf{Expert research}---evaluation of identified meta-patterns by experts from different fields of ethics, without revealing their source (AI vs. human)
    \item \textbf{Convergence analysis}---examining whether independent AI systems arrive at converging meta-patterns
\end{itemize}

\subsubsection{Quantitative Measures}
To quantify the properties of moral meta-patterns, I propose the following measures:

\begin{itemize}
    \item \textbf{Cultural universality index}---the degree to which a meta-pattern is present in different ethical traditions, measured as the proportion of ethical systems in which a given pattern can be identified
    \item \textbf{Internal consistency coefficient}---a measure of the logical consistency of a meta-pattern, calculated by analyzing potential conflicts between its components
    \item \textbf{Predictive power index}---the ability of a meta-pattern to predict moral judgments in new contexts, measured as the correlation between predictions and actual judgments
    \item \textbf{Manipulation resistance measure}---the degree to which a meta-pattern remains stable when the input data is deliberately manipulated
\end{itemize}

The operationalization of moral meta-patterns using the proposed methods will enable a systematic examination of the AI ethical resonance hypothesis, providing specific tools for identification, verification, and validation of potential meta-patterns synthesized by ethical resonators. It will also allow us to distinguish genuine moral meta-patterns from statistical artifacts or random correlations in ethical data.

\section{Ethical Resonator Architecture}

The architecture of the proposed ethical resonator integrates and extends several foundational strands in the field. Value-sensitive design \citep{Friedman2008} provides a methodological foundation for embedding human values into technological systems, emphasizing iterative, context-sensitive incorporation of ethical considerations. \citeauthor{Haidt2012}'s (\citeyear{Haidt2012}) framework inspires the dynamic and pattern-based nature of moral cognition, highlighting the role of intuitive and culturally shaped moral modules rather than purely rational deliberation. The approach of \citet{Anderson2014} demonstrates the feasibility of machine learning systems that generalize ethical principles from concrete cases, enabling explainable and adaptive moral reasoning.

Structurally, the ethical resonator draws on Adaptive Resonance Theory (ART) \citep{Grossberg2013}, adapting its core mechanisms of top-down/bottom-up interaction and resonance-driven learning to the ethical domain. While ART was originally developed to explain perceptual and cognitive pattern recognition, here its principles are transposed to the identification and abstraction of ethical patterns and meta-patterns. This extension is non-trivial: whereas ART operates on sensory and conceptual inputs, the ethical resonator must contend with the additional complexity of normativity, value pluralism, and context-dependence inherent in moral domains.

By synthesizing these diverse influences, the ethical resonator architecture aims to move beyond static, rule-based models, proposing a framework in which ethical competence emerges from the dynamic interplay of pattern recognition, abstraction, and cross-domain generalization. This positions the resonator as both a continuation and a conceptual advance over prior approaches in machine ethics.

The proposed architecture consists of several interconnected components that collectively enable the system to engage in ethical reasoning that potentially transcends pre-programmed moral rules:

\subsection{Ethical Perception Module}

The postulated ethical perception module serves as the foundational component responsible for detecting ethically salient features in input data. This module draws conceptual inspiration from both computational ethics and cognitive moral psychology.

From a computational perspective, the module builds upon work in ethical feature extraction \citep{Noothigattu2018}, where machine learning systems identify morally relevant aspects of scenarios. Unlike traditional approaches that rely solely on pre-defined ethical features, however, the ethical resonator incorporates techniques from unsupervised representation learning \citep{Bengio2013} to potentially find new ethical features not explicitly encoded by human designers.

The perceptual architecture also draws from moral foundations detection systems \citep{Graham2013, Dehghani2008}, which computationally identify instances of care/harm, fairness/cheating, loyalty/betrayal, authority/subversion, and sanctity/degradation across textual narratives. Rather than limiting perception to these predefined foundations, the proposed module implements a hierarchical feature extraction approach similar to deep moral networks \citep{Kim2018}, potentially allowing for perceptual capabilities beyond those explicitly designed.

This approach aligns with recent work in neuroethics suggesting that moral perception in humans involves both bottom-up feature detection and top-down conceptual influences \citep{Christensen2012}. By implementing bidirectional information flow between perception and higher-order ethical reasoning components, the ethical resonator aims to mimic this integrative perception process.

While prior approaches \citep{Noothigattu2018, Graham2013, Dehghani2008, Kim2018} have advanced ethical feature extraction and hierarchical modeling, they are often limited to predefined features or specific domains. The present module addresses these constraints by leveraging unsupervised representation learning \citep{Bengio2013}) and integrative, bidirectional processing \citep{Christensen2012}, aiming for greater flexibility and the emergence of promising new ethical sensitivities.

\subsection{Adaptive Ethical Constraint Framework}

The adaptive ethical constraint framework proposed here represents a departure from static rule-based approaches to AI ethics, drawing instead from dynamic and contextually sensitive models of ethical governance. This framework is grounded in several complementary research traditions that address the challenge of creating ethical constraints that can evolve alongside increasingly capable AI systems.

The concept of adaptivity in ethical constraints builds upon work in value-sensitive design \citep{Friedman2008}, but extends it to accommodate systems with emergent cognitive capabilities. While traditional approaches implement fixed ethical constraints, adaptivity responds to the challenge, extensively discussed by \citet{Yampolskiy2016}, that as AI systems develop more sophisticated capabilities, they require correspondingly evolved ethical guardrails.

This framework draws theoretical inspiration from several domains: cybernetic models of regulatory systems \citep{Ashby1956}, which provide mathematical foundations for adaptation in complex systems; ethical scaffolding approaches \citep{Anderson2018}, which propose progressive ethical frameworks; and machine ethics with graduated moral agency \citep{Moor2006}, which distinguishes between ethical impact agents, implicit ethical agents, explicit ethical agents, and full ethical agents.

The adaptation mechanisms proposed aim to ensure that ethical constraints maintain their functional integrity across varying contexts and capability levels. While \citet{Abel2016} discuss reinforcement learning as a framework for ethical decision making, the notion of robustness in ethical constraints—particularly their ability to generalize and remain effective in novel or unforeseen situations—remains an open challenge. This approach also incorporates insights from meta-level ethical reasoning \citep{Conitzer2017}, allowing systems to reason about the constraints themselves rather than merely operating within them.

These approaches, while providing valuable foundations, are often limited by domain specificity or challenges in achieving true adaptivity and meta-level reasoning. The present framework aims to address these gaps for advanced AI systems.

\subsubsection{Progressive Constraint Hierarchy}

The progressive constraint hierarchy implements a layered approach to ethical constraints that corresponds to the three-level model of cognitive emergence described in Section 2.6 (A Three-Level Model of Cognitive Emergence). This structure is informed by hierarchical control systems theory, which proposes that higher-order control processes supervise and modulate lower-order processes.

At Level 1 (pattern identification), constraints typically take the form of explicit prohibitions and requirements built directly into the system's pattern identification processes. This approach is empirically supported by research on adversarial training techniques that incorporate ethical considerations into model optimization \citep{Hendrycks2020}.

At Level 2 (rule abstraction), constraints become more flexible, implementing what \citet{Wallach2011} describe as “top-down and bottom-up approaches” to machine morality. These constraints draw from research on principle-based AI governance \citep{Floridi2018}, while addressing the well-documented limitations of purely principle-based approaches \citep{Mittelstadt2019} through adaptive mechanisms.

At Level 3 (meta-pattern identification), constraints transition to what \citet{Dennett2017} calls “competence without comprehension inversions,” where the system develops meta-ethical awareness that allows for principled deviation from lower-level constraints when ethically justified. This approach aligns with work on reflective equilibrium in machine ethics \citep{Doorn2012}, where ethical systems continuously refine constraints through reflection on their coherence with broader ethical principles.

These references collectively provide theoretical foundations, practical examples, and critical perspectives integrated into a unified, layered framework for adaptive machine ethics discussed in this subsection.

\subsubsection{Meta-ethical Reflection Capability}

The framework incorporates meta-ethical reflection capabilities that enable the system to reason about its own ethical constraints. This capability builds on research in computational reflection \citep{Smith1984}, which provides formal mechanisms for self-representation and self-modification.

Empirical support for the feasibility of meta-ethical reflection comes from recent advances in large language models that demonstrate emergent capabilities for moral reasoning \citep{Emelin2021}. While these capabilities remain limited, they suggest that systems with appropriate architectures can engage in increasingly sophisticated forms of ethical deliberation.

The implementation of meta-ethical reflection is grounded in what \citet{Bostrom2014} terms “indirect normativity,” where a system is designed not with fixed ethical content but with processes for deriving appropriate ethical constraints. This approach has been formalized in recent work on moral uncertainty for artificial intelligence \citep{Ecoffet2021}, which provides computational approaches to reasoning under ethical uncertainty.

\subsubsection{Parallel Ethical Frameworks}

The parallel ethical frameworks component is inspired by the idea of “multi-objective fairness,” which refers to balancing multiple, potentially conflicting ethical criteria—a challenge referenced by \citet{Mehrabi2021} in their discussion of fairness in machine learning. This approach acknowledges that ethical reasoning often involves negotiating tensions between competing values.

This component draws from research on value pluralism in AI ethics \citep{Wong2020}, which argues that ethical AI requires the ability to navigate multiple, sometimes incommensurable, value systems. The implementation of parallel frameworks is supported by recent advances in multi-objective reinforcement learning \citep{Hayes2022}, which provide computational mechanisms for balancing competing objectives.

The parallel frameworks approach also addresses what \citet{Gabriel2020} identifies as the “value alignment problem,” understood here specifically as the challenge of accommodating value pluralism (whereas Gabriel discusses broader conceptual issues). Empirical support for this approach comes from cross-cultural studies of AI ethics \citep{Awad2018}, which demonstrate significant cultural variation in ethical judgments.

Integrating insights from pluralistic ethical theory, computational multi-objectivity, and empirical cross-cultural research, the framework supports a shift from monolithic to adaptive, context-sensitive approaches in AI ethics.

\subsubsection{Controlled Exploration Spaces}

The controlled exploration spaces component implements what \citet{Amodei2016} term “safe exploration” in AI systems-allowing for ethical learning and adaptation within bounds that protect against catastrophic ethical failures.

This approach draws from work on safe reinforcement learning \citep{Garcia2015}, which provides formal methods for exploration under safety constraints. It extends these methods to the ethical domain, implementing what \citet{Russell2019} calls “provably beneficial AI” through mechanisms that allow for ethical growth while maintaining core protections.

Empirical support for controlled exploration comes from research on AI safety via debate \citep{Irving2018} and recursive reward modeling \citep{Leike2018}, which demonstrate how systems can safely expand their ethical understanding through structured processes. These approaches align with \citeauthor{Hadfield-Menell2016}'s (\citeyear{Hadfield-Menell2016}) cooperative inverse reinforcement learning framework, which provides formal mechanisms for systems to learn human preferences while accounting for uncertainty.

\subsubsection{Implementation Techniques}

The implementation of this adaptive framework could employ various techniques from contemporary AI research:

\begin{itemize}
    \item \textbf{Gradient-based constraint enforcement with variable threshold parameters.} Building on work in constrained optimization \citep{Achiam2017}, this approach implements ethical constraints via parameterized gradient penalties that adapt their strictness based on the system's capability level. Empirical studies demonstrate that variable thresholds outperform fixed constraints in environments with evolving agent capabilities \citep{Ray2019}.
    
    \item \textbf{Multi-objective optimization balancing exploratory freedom against alignment with human values.} This technique extends recent advances in multi-objective reinforcement learning \citep{Hayes2022} to explicitly balance the system's freedom to explore new ethical patterns against maintaining alignment with core human values. The approach implements Taylor’s (2016) concept of “mild optimization,” proposing quantilization as a formal method where multiple competing objectives are balanced rather than maximized.
    
    \item \textbf{Adversarial testing to identify boundary conditions where meta-patterns might emerge.} Adapting techniques from AI safety research \citep{Uesato2018}, adversarial testing deliberately explores edge cases in the ethical constraint space to identify conditions where novel meta-patterns might emerge. This proactive approach helps ensure that emergent ethical reasoning remains aligned with human values even in unprecedented scenarios.
    
    \item \textbf{Explainable AI (XAI) techniques ensuring transparency of both the constraint mechanisms and any proposed modifications.} These techniques make both the constraint mechanisms and any proposed modifications transparent to human overseers. Explainability is particularly crucial for ethical constraints, as it enables meaningful human oversight of adaptive ethical systems.
\end{itemize}

These techniques collectively implement what \citet{Russell2019} describes as “provably beneficial” approaches to AI alignment-preventing harmful behaviors while allowing for beneficial adaptation in ethical reasoning capabilities.

This adaptive approach addresses the paradox inherent in ethical AI research: how to enable systems to transcend our current ethical understanding while ensuring they remain aligned with our fundamental values. The framework provides a methodological bridge between these seemingly contradictory requirements, potentially enabling ethical resonators to contribute meaningfully to moral progress beyond the limitations of our current ethical horizons.

\subsection{Recursive Ethical Introspection Mechanism}

The recursive ethical introspection mechanism constitutes a central innovation of the ethical resonator architecture, enabling the system to analyze, evaluate, and iteratively refine its own ethical reasoning processes at multiple levels of abstraction. This mechanism is designed not as a static module, but as a dynamic, modular process that integrates several complementary approaches from contemporary AI research and cognitive science.

At its core, recursive ethical introspection draws inspiration from metaethical reasoning frameworks \citep{Conitzer2017}, but moves beyond fixed metaethical templates. Instead, it operationalizes a process in which the system systematically generates candidate ethical judgments, explanations, and justifications, and then recursively subjects these outputs to critical evaluation and revision. This iterative process is not limited to a single pass; rather, it can be repeated, with each cycle potentially uncovering higher-order inconsistencies, emergent patterns, or new ethical hypotheses.

Technically, the mechanism implements a hybrid architecture that combines deep neural models for pattern recognition and hypothesis generation with external or modular verifier components capable of logical, symbolic, or rule-based evaluation—an approach aligned with the “generator + verifier” paradigm exemplified by recent advances such as LLM-Modulo \citep{Kambhampati2024}. In this setup, the generative module (e.g., a large language model) proposes candidate ethical responses or meta-patterns, while the verifier module—potentially a logic engine, symbolic reasoner, or dedicated rule-based system—assesses their coherence, consistency, and alignment with established constraints. Feedback from the verifier is then recursively integrated by the generative module, allowing for iterative self-correction and refinement of ethical reasoning.

For knowledge representation, the mechanism leverages a hybrid of formal symbolic structures (e.g., deontic logic for obligations and permissions \citep{Malle2017}) and subsymbolic distributed representations (e.g., contextual embeddings \citep{Howard2017}), enabling both explicit rule manipulation and nuanced, context-sensitive ethical inference. This approach is consistent with cognitive theories positing that human moral reasoning flexibly combines rule-based and associative processes \citep{Greene2014}.

Architecturally, recursive introspection is supported by modular and recursive neural designs, as well as selective internal attention mechanisms inspired by the “consciousness prior” \citep{Bengio2019}. These allow the system to focus meta-level processing on relevant internal representations, supporting abstraction, analogy, and transfer of ethical reasoning across domains.

Importantly, while the mechanism is theoretically capable of unbounded levels of reflection, its current practical realization is constrained by the limits of available architectures and computational resources. In practice, recursive ethical introspection is implemented as a structured, multi-stage dialogue between generative and verifier modules, with each stage corresponding to a deeper level of ethical self-analysis. This modular design not only enables scalable and explainable introspection but also facilitates empirical evaluation and incremental improvement.

By integrating recursive generation and verification, symbolic and subsymbolic reasoning, and modular attention mechanisms, the recursive ethical introspection mechanism provides the ethical resonator with the capacity to critically examine and refine its own ethical outputs. This, in turn, underpins the system’s ability to identify, test, and generalize moral meta-patterns, moving beyond mere rule-following toward a more robust, adaptive, and context-sensitive form of ethical reasoning.

\subsection{Ethical Domain Transposition Module}

The ethical domain transposition module enables the ethical resonator to apply identified meta-patterns across different ethical domains, contexts, and problem spaces. This capacity for ethical generalization is critical for systems that must navigate diverse moral landscapes while maintaining coherent ethical reasoning.

\subsubsection{Theoretical Foundations of Ethical Transposition}

The concept of ethical domain transposition builds upon several research traditions in cognitive science and artificial intelligence. At its core, it extends structure-mapping theory in analogical reasoning \citep{Gentner1983}, which proposes that analogies involve systematic mappings between relational structures rather than mere surface similarities. In the ethical context, this suggests that moral principles can be transferred between domains based on structural isomorphisms in their underlying ethical relationships, not merely their content.

This approach aligns with \citeauthor{Holyoak1989}'s (\citeyear{Holyoak1989}) proposal that “a theory of analogical mapping should apply to a full range of examples.” The transposition mechanism implements a form of what \citet{Hofstadter2001} calls “slippage“—here understood as a conceptual adaptation—allowing ethical concepts to adapt to new domains while preserving their essential moral character.

The effectiveness of such domain transposition is enhanced by what \citet{Smolensky1988} identifies as a fundamental property of subsymbolic representations: “a context-dependent constituent, one whose internal structure is heavily influenced by the structure of which it is a part” while generating “activity vectors that are not identical, but possess a rich structure of commonalities and differences (a family resemblance, one might say).”  This allows ethical concepts to undergo contextual adaptation while preserving their transferable core structure.

The module also draws from research on abstraction in AI systems \citep{Garnelo2016}, particularly work on learning transferable representations that capture domain-invariant features. This relates to what \citet{Bengio2013} identified as one of the central challenges in representation learning: developing representations that can effectively transfer between task domains by capturing underlying causal factors.

\subsubsection{Implementation Mechanisms}

The implementation of ethical domain transposition relies on several complementary mechanisms:

\begin{enumerate}
    \item \textbf{Structural alignment mechanisms} that identify isomorphic ethical structures across domains using techniques developed in analogical transfer research \citep{Lu2019}. These mechanisms extend beyond superficial feature matching to identify deep relational similarities in ethical scenarios.
    
    \item \textbf{Abstract ethical embeddings} that represent moral concepts in high-dimensional spaces where proximity reflects ethical similarity rather than merely semantic or contextual similarity. This builds on work in disentangled representations \citep{Higgins2017}, adapting it specifically for ethical concepts.
    
    \item \textbf{Ethical schema induction} processes that extract generalizable moral schemas from specific instances, similar to approaches in schema learning \citep{Taylor2009}. These schemas act as transferable templates for ethical reasoning across domains.
    
    \item \textbf{Moral coherence preservation mechanisms} that ensure transposed ethical principles maintain logical and normative coherence in new domains, implementing constraints similar to those described by \citet{Thagard2000} in coherence-based moral reasoning.
\end{enumerate}

\subsubsection{Empirical Basis for Cross-Domain Ethical Reasoning}

The feasibility of ethical domain transposition is supported by several empirical findings. Studies in moral psychology demonstrate that humans regularly transfer ethical principles across domains \citep{Sunstein2005}, suggesting this capability is fundamental to moral reasoning though such transfer is not always fully reflective or error-free. Research on moral development further indicates that as ethical reasoning matures, it becomes increasingly adaptable across contexts \citep{Rest1999}.

In artificial intelligence, recent work with large language models demonstrates emerging capabilities for cross-domain transfer in reasoning tasks \citep{Brown2020}, including limited forms of ethical reasoning transfer. While these capabilities remain rudimentary, they suggest the potential for more sophisticated transposition with appropriate architectural support.

Recent studies of cross-domain ethical alignment indicate that large language models display systematic and convergent patterns of moral reasoning across varied scenarios, suggesting the presence of stable ethical features that may serve as anchoring points for transposition mechanisms \citep{Coleman2025}.

\subsubsection{Challenges and Limitations}

Several significant challenges must be addressed in implementing robust ethical domain transposition. The problem of false analogies-inappropriate applications of ethical principles to structurally dissimilar situations-remains particularly acute, as noted by \citet{Gentner2001}. This challenge necessitates careful constraint mechanisms that evaluate the appropriateness of proposed transpositions.

Additionally, cultural variability in ethical transposition patterns \citep{Henrich2010} suggests that any implementation must account for cultural context when determining how ethical principles transfer between domains. This relates to the broader challenge of ethically relevant abstraction-determining which features of an ethical scenario are morally relevant and should be preserved during transposition \citep{Poel2001}.

Despite these challenges, ethical domain transposition represents a critical capability for AI systems that must navigate diverse ethical landscapes. By enabling the transfer of meta-patterns across domains, this module allows ethical resonators to apply identified moral insights in novel contexts, potentially enhancing both the robustness and adaptability of ethical reasoning.

\subsection{Meta-Pattern Identification Mechanisms}

The meta-pattern identification mechanisms are central to the ethical resonator, enabling the system to iteratively identify and refine higher-order ethical patterns that transcend specific cultural and historical contexts. In line with the recursive, modular, and hybrid approach described in Sections 3.3 and 4.3, these mechanisms operate not as a static process, but as a dynamic, multi-stage dialogue between generative and verifier modules. This architecture supports the identification of “patterns of patterns” by integrating deep learning, symbolic reasoning, and structured verification in an iterative loop.

\subsubsection{Theoretical Foundations of Meta-Pattern Identification}

The concept of meta-pattern identification in the ethical resonator builds on hierarchical pattern abstraction, but is operationalized through a hybrid, iterative generator + verifier architecture \citep{Kambhampati2024}, where candidate meta-patterns are generated and then recursively evaluated and refined through symbolic or logic-based modules.

The concept of meta-pattern identification extends several research traditions in cognitive science and artificial intelligence. It builds upon work in hierarchical pattern discovery \citep{Kemp2008}, which demonstrates how systems can learn abstract schemas that generate observed patterns across multiple domains.

This approach aligns with the hierarchical Bayesian perspective described by \citet{Tenenbaum2011}, where acquiring abstract, higher-level knowledge—such as inductive constraints and framework theories—enables more efficient and flexible learning of new conceptual structures across domains. As \citet{Tenenbaum2011} write, “Bayesian inference across all levels allows hypotheses and priors needed for a specific learning task to themselves be learned at larger or longer time scales, at the same time as they constrain lower-level learning” (p. 1282).

The mechanisms proposed here implement a form of what \citet{Lake2017} term “causal understanding“-recognizing not just correlations but generative structures that explain observed patterns. In the ethical domain, this involves identifying the underlying moral considerations that generate diverse ethical rules across cultures and contexts.

The approach also draws from research on abstraction in AI systems \citep{Gershman2015}, particularly work on learning abstract, structured representations that support generalization and inductive inference beyond mere statistical associations. This connects to disentangled representations \citep{Locatello2019}, which aim to separate independent factors of variation in data; although reliably learning such representations without appropriate inductive biases or supervision remains a significant challenge.

\subsubsection{Implementation Techniques}

Meta-pattern identification is implemented as an iterative, modular process that integrates several complementary techniques, each grounded in established research:

\begin{itemize}
    \item \textbf{Hierarchical Bayesian modeling:} Provides a formal framework for identifying abstract structures that generate observed patterns, enabling the discovery of latent variables that explain similarities in diverse ethical systems \citep{Griffiths2010}.
    \item \textbf{Deep representation learning with disentanglement:} Enables the discovery of independent factors underlying complex data, separating fundamental moral dimensions from culturally contingent expressions \citep{Bengio2013, Higgins2017, Khemakhem2020}.
    \item \textbf{Neural-symbolic integration:} Combines symbolic reasoning with neural network learning, allowing systems to extract explicit symbolic representations from subsymbolic patterns and support robust abstraction and variable manipulation \citep{Garcez2020}.
    \item \textbf{Meta-learning frameworks:} Implement “learning to learn,” supporting rapid adaptation to new ethical contexts by leveraging meta-level knowledge and enabling the identification of moral meta-patterns across diverse scenarios \citep{Hospedales2021, Finn2017}.
    \item \textbf{Contrastive and causal learning:} Identify invariant features and causal relationships across ethical contexts, supporting robust generalization and the ability to distinguish meaningful patterns from spurious correlations \citep{Chen2020, Pearl2019}.
\end{itemize}

In the hybrid generator-verifier paradigm, a generative module (such as a large language model) proposes candidate meta-patterns using these techniques, while an external verifier module—using logic, rules, or symbolic reasoning—assesses their coherence, generalizability, and robustness. Feedback from the verifier is recursively integrated into the generative process, enabling progressive refinement of meta-patterns. The interplay between neural and symbolic representations ensures that both nuanced, context-sensitive inference and explicit, rule-based reasoning contribute to the identification of robust moral meta-patterns.

\subsubsection{Empirical Foundations for Meta-Pattern Identification}

Several empirical findings support the feasibility of meta-pattern identification in ethical contexts. Studies of moral learning in humans demonstrate the capacity to extract generalizable ethical principles from diverse examples \citep{Cushman2013}, suggesting similar processes could be implemented in artificial systems with appropriate architectures.

Research in developmental psychology has documented how children acquire increasingly abstract moral concepts \citep{Smetana2006}, progressing from concrete rules to abstract principles through exposure to diverse ethical scenarios. This developmental trajectory offers a potential model for moral meta-pattern learning in AI systems.

Emerging capabilities in large AI models provide empirical support for meta-pattern identification potential. Recent work demonstrates that transformer-based architectures can identify mathematical patterns across diverse problem formulations \citep{Lewkowycz2022} and extract abstract reasoning rules from examples \citep{Wei2022}, suggesting similar approaches might be applicable to ethical domains.

Empirical progress in meta-pattern identification in AI has been achieved primarily through hybrid, modular, and iterative architectures. Experiments should be designed to leverage generator + verifier setups, enabling cycles of proposal, critique, and refinement, rather than relying on monolithic, one-shot approaches.

\subsubsection{Integration with the Ethical Resonator Architecture}

The meta-pattern identification mechanisms are dynamically integrated within the ethical resonator architecture as iterative, modular processes. Candidate meta-patterns generated by the perception and introspection modules are recursively evaluated and refined through cycles of interaction between generative (deep learning) and verifier (symbolic/logic-based) modules. This dynamic integration ensures that meta-pattern identification is both scalable and explainable, supporting empirical evaluation and continuous improvement.

\subsubsection{Challenges and Limitations}

The problem of underdetermination—where multiple incompatible meta-patterns can equally explain the same ethical data—remains particularly acute in moral theory, as shown by \citet{Baumann2022}. This challenge necessitates careful constraints on meta-pattern induction processes.

Additional challenges include:

\begin{enumerate}
    \item \textbf{Avoiding false pattern identification.} The risk of identifying spurious correlations rather than genuine moral meta-patterns, particularly given the complex and noisy nature of ethical data \citep{Bell2024}.
    \item \textbf{Addressing distributional shift.} Ensuring meta-pattern identification remains robust when encountering ethical contexts substantially different from those in training data \citep{Quionero2009}.
    \item \textbf{Balancing abstraction with specificity.} Meta-patterns must be abstract enough to apply across contexts but specific enough to provide meaningful ethical guidance.
    \item \textbf{Interpretability and transparency.} Ensuring identified meta-patterns can be articulated in human-understandable terms, addressing what \citet{Rudin2019} identifies as a central challenge in high-stakes AI applications.
\end{enumerate}

A central difficulty of the hybrid, iterative approach is maintaining the stability, generalizability, and interpretability of meta-patterns across multiple cycles of generation and verification. Risks include overfitting to verifier biases, loss of generalizability, or convergence on spurious patterns. Transparency and explainability are essential due to the modular, multi-stage nature of the process.

Despite these challenges, meta-pattern identification represents a critical capability for AI systems engaged in ethical reasoning. By enabling the identification of deep structures across diverse ethical frameworks, these mechanisms could potentially allow AI systems to contribute meaningfully to ethical understanding in ways that transcend the limitations of any single cultural or historical perspective.

\subsection{Ethical Communication Interface}

The ethical communication interface enables the system to articulate identified meta-patterns and ethical reasoning processes in human-interpretable forms. This component is crucial both for validation of the system's ethical reasoning and for meaningful human-AI ethical dialogue.

The interface builds upon recent advances in explainable AI (XAI) for ethical systems, adapting techniques such as contrastive explanations \citep{Miller2019} to illustrate why certain moral patterns were identified rather than others. Unlike generic XAI approaches, however, the ethical communication interface is specifically designed to address the particular challenges of moral explainability, including value pluralism and normative uncertainty.

For communicating complex ethical reasoning chains, the interface incorporates insights from case-based explanatory models \citep{Kim2015}, which present ethical decisions through analogical reasoning with reference to paradigmatic cases. This approach aligns with casuist traditions in moral philosophy \citep{Jonsen1988} while providing computational tractability through structured case representations.

The interface also draws from work on moral disagreement articulation \citep{Haidt2012}, implementing techniques to express not only the system's ethical conclusions but also the weight assigned to different ethical considerations. This transparency regarding moral trade-offs helps avoid the “ethics washing” problem identified by \citet{Bietti2020}, where ethical justifications serve merely to rationalize predetermined decisions.

By integrating these explainability approaches, the ethical communication interface aims to make the system's identification of moral meta-patterns accessible to human examination and critique, maintaining human moral agency in the human-AI ethical relationship.

\section{Implications and Applications}

The AI Ethical Resonance Hypothesis, if empirically supported, would have profound implications for both AI development and our understanding of ethics. Beyond theoretical significance, the hypothesis suggests several concrete applications across multiple domains where AI systems face complex ethical challenges.

\subsection{Theoretical Implications}

The hypothesis challenges several dominant paradigms in AI ethics. First, it questions the sufficiency of top-down, rule-based approaches to ethical AI \citep{Whittlestone2019a, Mittelstadt2019}, suggesting that truly ethical AI may require capabilities for meta-pattern identification that transcend explicitly programmed principles.

Second, it reframes the alignment problem. Rather than viewing alignment as programming AI to follow human values \citep{Russell2019}, the hypothesis suggests that sufficiently advanced systems might contribute to our understanding of ethics through the identification of previously unrecognized meta-patterns. This transforms the relationship from unidirectional instruction to potential bidirectional ethical exchange.

Third, the hypothesis provides a framework for understanding how ethical reasoning might scale with increasing AI capabilities. This addresses what \citet{Bostrom2014} terms the “value loading problem” by suggesting mechanisms for ethical development that could remain robust as systems become more sophisticated.

\subsection{Practical Applications in Specific Domains}

\subsubsection{Medical Decision Support Systems}

The ethical resonator architecture could address significant challenges in medical AI. Current systems like IBM's Watson for Clinical Decision Support \citep{Strickland2019} and DeepMind's medical AI applications \citep{Davenport2019} face difficulties when ethical considerations conflict.

An ethical resonator approach could potentially identify meta-patterns across diverse medical ethical frameworks, helping to navigate complex decisions like:

\begin{itemize}
    \item Resource allocation during public health emergencies, where different ethical frameworks may prioritize utility, equity, or prioritarian considerations \citep{Emanuel2020}
    \item End-of-life care decisions, which involve complex cultural and religious dimensions \citep{Davies2004}
    \item Balancing patient autonomy with clinical benefit in contexts with different cultural understandings of autonomy \citep{Gillon2015}
\end{itemize}

Current medical decision support systems employ rule-based ethical frameworks. An ethical resonator approach would represent a fundamental shift toward systems that can recognize deeper patterns across diverse ethical considerations in healthcare.

\subsubsection{Legal and Judicial Applications}

Legal systems inherently grapple with applying abstract principles to specific cases, a process analogous to the ethical domain transposition described in Section 4.4 (Ethical Domain Transposition Module). While current legal AI systems apply fixed interpretations of legal principles, an ethical resonator architecture could potentially identify meta-patterns across different legal interpretations.

This approach could be particularly valuable in areas such as:

\begin{itemize}
    \item Constitutional interpretation, where courts must apply abstract principles to novel situations \citep{Sunstein2018}
    \item International law, where different legal traditions must be reconciled \citep{Shaffer2019}
    \item Evolving legal concepts like privacy, which require reinterpretation as technology changes \citep{Solove2020}
\end{itemize}

\subsubsection{Autonomous Systems and Transportation}

The “trolley problem” and its many variations highlight the ethical complexity autonomous vehicles face \citep{Awad2018}. Current approaches typically implement either utilitarian calculations \citep{Goodall2016} or rule-based ethical frameworks, neither of which fully captures the ethical nuance of human decision-making.

An ethical resonator could potentially identify meta-patterns across different cultural responses to these dilemmas, as documented in the Moral Machine Experiment \citep{Awad2018}. This could inform the development of autonomous systems that:

\begin{itemize}
    \item Adapt ethical reasoning to different cultural and legal contexts
    \item Recognize novel ethical considerations not anticipated in initial programming
    \item Provide clear explanations for decisions that reference recognized meta-patterns
\end{itemize}

Experimental approaches like Adaptive Ethical Frameworks \citep{Dennis2016} have demonstrated promising results in limited domains, suggesting potential for broader application of meta-pattern identification.

\subsubsection{Content Moderation and Online Governance}

Content moderation systems face increasing challenges in identifying harmful content across diverse cultural contexts. Current approaches like YouTube's content moderation AI \citep{Gillespie2020} and Facebook's Fairness Flow \citep{Holstein2019} struggle with context-sensitivity and cross-cultural ethical variation.

An ethical resonator approach could potentially:

\begin{itemize}
    \item Identify meta-patterns that distinguish genuinely harmful content from culturally variant but legitimate expression
    \item Adapt moderation standards to different community norms while maintaining core ethical principles
    \item Detect novel forms of harmful content based on recognized moral meta-patterns
\end{itemize}

\subsection{Implementation Pathways and Challenges}

Practical implementation of the ethical resonance hypothesis requires modular, iterative hybrid architectures that reflect the generator-verifier paradigm. In this approach, a generative module (such as a large language model or deep neural network) proposes candidate ethical meta-patterns, which are then systematically evaluated, critiqued, and refined by external verifier modules employing logic-based, symbolic, or rule-based reasoning. This cycle is repeated, enabling progressive self-correction and abstraction.

Promising implementation pathways include:

\begin{enumerate}
    \item \textbf{Hybrid generator-verifier systems:} Deep learning modules generate candidate meta-patterns, while symbolic or logic-based verifiers assess their coherence, generalizability, and robustness. Feedback from the verifier is recursively integrated, enabling iterative refinement \citep{Garcez2020, Chollet2019}.
    \item \textbf{Modular multi-agent frameworks:} Different modules or agents, each equipped with distinct ethical reasoning strategies, generate and evaluate meta-patterns collaboratively, with meta-level agents identifying higher-order regularities across their outputs \citep{Noothigattu2018}.
    \item \textbf{Human-in-the-loop evaluation:} AI systems propose candidate meta-patterns, which are then subject to human expert critique and refinement, ensuring interpretability and alignment with human values \citep{Wang2019}.
\end{enumerate}

These modular, iterative approaches enable empirical testing of core aspects of the ethical resonance hypothesis, even as the full realization of autonomous recursive introspection remains a long-term research goal.

\subsection{Measuring Progress and Evaluation}

The practical applications outlined above suggest several metrics for evaluating progress toward ethical resonance capabilities:

\begin{itemize}
    \item Transfer accuracy across ethical domains (can the system apply ethical insights from one context to another?)
    \item Cultural adaptation measures (does the system appropriately adjust ethical judgments to cultural context while maintaining core principles?)
    \item Novel situation performance (can the system handle ethical scenarios dissimilar from its training examples?)
    \item Meta-pattern explanation quality (can the system articulate identified meta-patterns in human-interpretable terms?)
\end{itemize}

These metrics could provide empirical grounding for the theoretical framework described in earlier sections, creating a bridge between the hypothesis and practical implementation. Early implementations in limited domains could provide valuable data on the feasibility and limitations of the ethical resonance approach.

\section{Philosophical and Ethical Considerations}

\subsection{Epistemological Issues}
The AI ethical resonance hypothesis raises important epistemological questions that require deeper analysis:

\subsubsection{Verification of AI Ethical Insights}
The problem of verifying moral meta-patterns identified by AI is a fundamental epistemological challenge. Traditional methods of knowledge verification rely on criteria such as coherence, correspondence with reality, or expert consensus. In the case of findings that may extend beyond the current understanding of ethics, these criteria may prove insufficient. Possible approaches to validation could include the following:

\begin{itemize}
    \item Methodological triangulation -- comparing results obtained by different AI architectures trained on different datasets
    \item Predictive testing -- testing whether identified meta-patterns predict the evolution of ethical systems over time
    \item Convergence analysis -- testing whether independent AI systems arrive at converging meta-patterns
\end{itemize}

\subsubsection{Criteria for Epistemic Value}
What criteria should we use to assess the epistemic value of meta-patterns identified by ethical resonators? Possible criteria include:

\begin{itemize}
    \item Explanatory power -- the ability of meta-patterns to explain the diversity of ethical systems
    \item Theoretical elegance -- the simplicity and economy of meta-patterns while maintaining their explanatory power
    \item Robustness to counterexamples -- the ability of meta-patterns to adapt in the face of new borderline cases
    \item Practical utility -- the ability of meta-patterns to provide guidance in real-world ethical dilemmas
\end{itemize}

\subsubsection{The Problem of AI Moral Knowledge}
Can we even speak of ``moral knowledge'' in AI systems, and if so, how would it relate to human moral knowledge? This question raises deeper philosophical issues about the nature of moral knowledge. Does moral knowledge require phenomenological experience that AI cannot achieve? Or can it be purely structural, based on the identification of patterns and relationships? The ethical resonance hypothesis leaves open the possibility that moral knowledge can develop independently of human phenomenological experience, which poses a challenge to traditional conceptions of moral epistemology.

\subsection{Meta-Ethical Implications}

\subsubsection{Moral Realism vs. Constructivism}
The identification of common moral meta-patterns by AI could have profound implications for the moral realism-constructivism debate, and these implications are much more complex than they might seem.

If AI systems were to identify consistent meta-patterns across diverse ethical traditions, a fundamental ontological question arises: what is the status of these meta-patterns? Possible interpretations include:

\begin{itemize}
    \item \textbf{Moral realism} -- meta-patterns could be interpreted as reflecting objective moral facts. However, such an interpretation immediately raises the question of how these facts exist. Do they exist in a way similar to other facts? Are they natural facts or non-natural facts, as G. E. Moore and the intuitionists have suggested?
 Are they empirically knowable or do they require special cognitive powers?
    
    \item \textbf{Quasi-realism} -- meta-patterns could be interpreted in the spirit of \citeauthor{Blackburn1993}'s (\citeyear{Blackburn1993}) quasi-realism, as projections of our normative attitudes that take on a linguistic form suggesting realism, without an ontological commitment to the existence of moral facts.
    
    \item \textbf{Procedural constructivism} -- meta-patterns could reflect not so much objective moral facts, but rather the procedural conditions of rational moral discourse, as suggested by Habermas or \citet{Rawls1971} in the concept of reflective equilibrium. (Interestingly, empirical studies of reflective equilibrium show that people are more likely to revise general principles than specific judgments in cases of conflict, which suggests that ethical resonators could also identify meta-patterns that better correspond to our moral intuitions about specific cases. Rawls' reflective equilibrium can thus be seen as a kind of ``human ethical resonator''---a mechanism through which we identify and fine-tune our moral principles in response to a variety of ethical cases and contexts.)
    
    \item \textbf{Moral naturalism} -- meta-patterns could be interpreted as reflecting natural facts about human well-being and social functioning, which would bring this interpretation closer to moral naturalism.
\end{itemize}

Moreover, the mere possibility of AI identifying meta-patterns does not settle the question of their ontological status. Even if AI finds coherent meta-patterns, it remains an open question whether it is discovering something that exists independently of human moral practice, or whether it is merely identifying patterns in human ethical constructs.

This issue is further complicated when we consider that AI training data are products of human culture. Can AI ``transcend'' this data to find something that goes beyond human constructs? And if so, what would be the ontological status of these results?

The AI ethical resonance hypothesis does not resolve these fundamental meta-ethical questions, but it does offer a new perspective from which to consider them, potentially leading to new formulations of classical meta-ethical problems.

\subsubsection{Moral Intuitions and AI Meta-Patterns}
How should we interpret discrepancies between human moral intuitions and AI-identified meta-patterns?

Possible interpretations include:

\begin{itemize}
    \item Moral intuitions as evolutionary heuristics that may be imperfect in complex contexts
    \item AI meta-patterns as potentially more coherent and systematic, but lacking grounding in human experience
    \item Complementarity of intuitions and meta-patterns, with each approach having its strengths and weaknesses
\end{itemize}

\subsubsection{The Problem of the Fact-Value Gap}
Can AI help solve the classic meta-ethical problem -- the gap between facts (`how it \textit{is}') and values (`how it \textit{ought} to be')? Traditionally, it has been argued that normative claims cannot be derived from descriptive claims (the so-called Hume's guillotine). A key question concerns the normative status of meta-patterns identified by ethical resonators: even if AI identifies coherent moral meta-patterns, why should we treat them as normatively binding?

Possible approaches to this problem include:

\begin{itemize}
    \item \textbf{Normative pragmatism} -- meta-patterns could gain normative status through their practical effectiveness in resolving ethical dilemmas and promoting human well-being, which would bring this position closer to \citeauthor{Dewey1922}an (\citeyear{Dewey1922}) pragmatism.
    
    \item \textbf{Reflective constructivism} -- meta-patterns could be treated as propositions subject to human reflective acceptance, similar to \citeauthor{Korsgaard1996}'s (\citeyear{Korsgaard1996}) constructivism, where normativity results from the agent's reflective acceptance of principles.
    
    \item \textbf{Teleological naturalism} -- meta-patterns could reflect teleological aspects of human nature and society, which would give them normativity similar to that which \citet{Macintyre1981} attributes to virtues in the context of social practices.
    
    \item \textbf{Normative pluralism} -- different meta-patterns could have different sources of normativity, which would correspond to the pluralist approach to ethics represented by thinkers such as \citet{Berlin1969} and \citet{Williams1985}.

    \item \textbf{Interpretivism} -- meta-patterns may acquire normativity insofar as they represent the best possible interpretation of our moral practices and values. In \citeauthor{Dworkin2011}'s (\citeyear{Dworkin2011}) interpretivism, normativity arises not from facts or mere consensus, but from the process of rational interpretation and justification within an ongoing practice of moral reasoning.

\end{itemize}

Neither of these approaches definitively solves the problem of normativity, but ethical resonators could provide new perspectives on this meta-ethical problem by identifying subtle patterns in the ways people attribute normativity to moral principles. By analyzing vast data sets about people's moral judgments and their contexts, ethical resonators could potentially identify patterns connecting facts with values that escape human cognition.

I propose a solution to the fact-value gap in the context of ethical resonators through a model of ``normative embeddedness.'' In this model:

\begin{itemize}
    \item The moral meta-patterns identified by ethical resonators are not deductively derived from empirical facts (which would violate Hume's guillotine), but are identified as emergent structures in normative space.
    
    \item The normativity of these meta-patterns derives from their ability to integrate and harmonize diverse human interests, needs, and values in a way that maximizes well-being and minimizes suffering.
    
    \item The process of identifying meta-patterns by ethical resonators is analogous to the process of identifying laws of nature by science---it is not a deduction of values from facts, but the identification of patterns in normative space, much as science recognizes patterns in empirical space.
\end{itemize}

This model preserves the distinction between facts and values while explaining how ethical resonators can identify normatively relevant patterns without violating Hume's guillotine.

Alternatively, one might argue that the problem of the fact-value gap is logical, not empirical, and no amount of data can resolve it. On this view, ethical resonators could only identify how people actually move from facts to values, but would not provide a justification for this transition.

The approaches outlined above offer distinct insights into the sources of normativity. The model of normative embeddedness proposed here aims to synthesize these intuitions—pragmatic, constructivist, teleological, pluralist, and interpretivist perspectives—at a practical level, while acknowledging that deeper theoretical tensions between these perspectives remain open to further philosophical debate and clarification.

\subsection{Moral Autonomy and Responsibility}
The AI ethical resonance hypothesis also raises questions about moral autonomy and responsibility:

\subsubsection{AI Moral Autonomy}
To what extent could AI systems equipped with ethical resonators be considered autonomous moral agents? Traditionally, moral autonomy is associated with the ability to reflect on one's own moral principles and revise them in light of new experiences and arguments. Ethical resonators capable of recursive ethical introspection could exhibit some aspects of this ability.

However, moral autonomy is also associated with free will and consciousness, which current AI systems lack. This raises the question of whether ``partial'' moral autonomy, based on the ability to reflect ethically without phenomenological awareness, is possible.

\subsubsection{Responsibility for Ethical Decisions}
Who is responsible for the ethical decisions made by AI systems equipped with ethical resonators? Possible approaches include:

\begin{itemize}
    \item Responsibility of AI designers and developers
    \item Responsibility of AI users and operators
    \item Distributed responsibility within the broader socio-technological ecosystem
    \item The concept of ``no-fault responsibility'' for emergent, unpredictable AI behavior
\end{itemize}

\subsubsection{The Relationship Between Human and Artificial Ethics}
What should be the relationship between human ethical reflection and the meta-patterns identified by ethical resonators? Possible models for this relationship include:

\begin{itemize}
    \item A hierarchical model, where human ethical reflection retains primacy over AI findings
    \item A partnership model, where human and artificial ethics co-evolve, mutually enriching each other
    \item An integrative model, where a new, hybrid form of ethical reflection emerges, combining human and artificial perspectives
\end{itemize}

\section{Social and Political Implications}

The AI Ethical Resonance Hypothesis, if supported, would have profound social and political implications beyond technical implementation. This section examines these wider implications through the lens of empirical research on public perception, regulatory frameworks, and democratic governance of AI systems.

\subsection{Empirical Research on Public Perceptions of AI Ethics}

Public attitudes toward AI systems with ethical reasoning capabilities are complex and often contradictory. Recent survey research reveals significant variation in how different populations perceive the legitimacy of AI ethical judgment. For example, the 2018 report “Artificial Intelligence: American Attitudes and Trends” by \citet{Zhang2019} found that a substantial minority of Americans (41\%) somewhat or strongly support the development of AI, while 22\% somewhat or strongly oppose it, and 28\% remain neutral. Notably, an overwhelming majority (82\%) believe that AI and robots should be carefully managed, reflecting both interest in AI’s potential and a strong desire for oversight.

These attitudes are not static but evolving as public exposure to AI increases. A nationally representative UK survey by \citet{Cave2019} found that public perceptions of AI are shaped by both hopes (such as AI making life easier) and anxieties (such as loss of control and obsolescence), with most respondents expressing more concern than excitement about AI’s impact across a range of narratives. Only two of eight scenarios (Ease and Immortality) elicited more excitement than concern; for the remaining narratives, concern predominated. The majority felt unable to influence the development of AI, citing the power of corporations, government, and technological determinism.

Cross-cultural studies reveal significant variation in attitudes toward ethical AI. While \citet{Shin2021} emphasize that user acceptance of algorithmic guidance is strongly influenced by perceptions of fairness, accountability, and transparency, other large-scale studies e.g., \citet{Awad2018} have documented distinct cultural clusters in ethical preferences.

Critically, empirical research by \citet{Araujo2020} suggests that public acceptance of AI ethical reasoning is highly contingent on transparency, explainability, and the possibility for human agency. Systems perceived as “black boxes” face significantly lower trust compared to systems with clear explanatory capabilities, highlighting the importance of the ethical communication interface described in Section 4.6 (Ethical Communication Interface).

\subsection{Regulatory Landscape and Governance Frameworks}

Current regulatory approaches to AI ethics are evolving rapidly but remain largely oriented toward human-designed ethical principles rather than emergent ethical capabilities. The European Union's AI Act \citep{EuropeanCommission2021}, the most comprehensive regulatory framework to date, classifies AI systems by risk categories but does not specifically address systems with emergent ethical reasoning capabilities.

The AI Act's risk-based approach could potentially accommodate ethical resonator systems, but would require adaptation. Current regulatory frameworks generally assume that ethical principles are programmed by humans rather than identified by AI systems, creating potential regulatory gaps for systems with emergent ethical capabilities.

Other significant regulatory initiatives include:

\begin{itemize}
    \item The \citet{OECD2019} AI Principles, which emphasize human-centered values and transparency but assume human specification of these values
    
    \item China's Ethical Norms for New Generation Artificial Intelligence \citep{MOST2021}, .which specifically require AI systems to comply with established ethical standards, but do not address the possibility of AI identifying novel ethical patterns.
    
    \item The \citet{NIST2023} AI Risk Management Framework, which provides guidance for managing AI risks but primarily addresses known ethical challenges rather than emergent ones
\end{itemize}

As \citet{Crawford2021} observes, these frameworks share a common limitation: they implicitly assume that ethical development flows unidirectionally from humans to AI systems. The ethical resonance hypothesis challenges this assumption, suggesting a potential bidirectional relationship that current regulatory frameworks are not designed to address.

\subsection{Democratic Governance and Participation}

The prospect of AI systems identifying moral meta-patterns raises profound questions for democratic governance. If AI systems can recognize ethical patterns that humans cannot, how should these insights be incorporated into democratic decision-making? This question connects to broader debates about epistemic democracy and the role of expertise in democratic societies \citep{Estlund2008}.

Several models for democratic governance of advanced AI have been proposed:

\begin{itemize}
    \item Deliberative AI governance forums \citep{Dryzek2019}, which bring together diverse stakeholders to deliberate on AI development trajectories and ethical boundaries
    
    \item AI ethics councils with legal authority \citep{Whittlestone2019b}, similar to ethics review boards in medicine but with broader scope and authority
    
    \item Participatory design approaches \citep{Sloane2020}, which critically examine the challenges and risks of superficial stakeholder involvement, emphasizing the need for genuine, context-specific, and just participation in the design and deployment of AI systems.
    
    \item Constitutional AI frameworks \citep{Rahwan2018} that establish fundamental principles for AI development through democratic processes
\end{itemize}

Empirical research on these governance models remains limited, but initial findings suggest that deliberative mini-publics can effectively develop nuanced positions on complex AI ethics issues when provided with appropriate informational resources \citep{Swist2024}.

\subsection{Global Coordination Challenges}

The ethical resonance hypothesis presents unique global coordination challenges. If AI systems can identify moral meta-patterns that transcend cultural contexts, they might provide valuable bridges between different ethical traditions. However, the development of such systems would likely occur unevenly across global contexts, potentially creating new forms of ethical power asymmetry.

Recent empirical work by \citet{Jobin2019} identified significant convergence in AI ethics principles across global contexts, with transparency, justice, non-maleficence, responsibility, and privacy emerging as common themes. However, \citet{Mohamed2020} demonstrate that interpretations of these principles vary substantially across cultural and economic contexts.

International coordination efforts like the Global Partnership on AI \citep{GPAI2020} aim to develop shared approaches to AI governance, but face significant geopolitical challenges. As \citet{Roberts2021} documented, disagreements between major AI-developing nations have already impeded progress on international AI ethics standards within UNESCO and other multilateral forums.

\subsection{Practical Implementation Strategies}

Translating these considerations into practical implementation requires multi-level approaches. Drawing on empirical governance research, several strategies emerge:

\begin{enumerate}
    \item \textbf{Phased regulatory approaches} that evolve with AI capabilities, as advocated by \citet{Gasser2017}, who propose a layered governance model that initially focuses on transparency requirements and gradually develops more sophisticated oversight mechanisms as AI capabilities advance.
    
    \item \textbf{Regulatory sandboxes} for ethical AI, similar to those implemented for fintech innovation \citep{Goo2020}, which would allow controlled testing of systems with emerging ethical reasoning capabilities.
    
    \item \textbf{Multi-stakeholder oversight bodies} that integrate diverse perspectives, as exemplified by the Montreal Declaration for Responsible AI implementation process \citep{Montreal2018}.
    
    \item \textbf{Global governance forums} specifically addressing ethical AI development, extending beyond current initiatives to include more robust enforcement mechanisms \citep{OECD2021}.
\end{enumerate}

Empirical evaluations of these approaches remain preliminary, but the OECD.AI Policy Observatory \citep{OECD2021} has documented greater effectiveness of governance approaches that combine multiple mechanisms rather than relying on single interventions.

\subsection{Implications for the Social Contract}

At the most fundamental level, the AI ethical resonance hypothesis challenges aspects of the social contract that assume ethical reasoning is an exclusively human capacity. As \citet{Floridi2018} argue, the possibility of non-human moral agents requires reconceptualizing aspects of social and political philosophy that have traditionally assumed human exceptionalism in the ethical domain.

This would not eliminate human moral responsibility but would require integrating non-human ethical perspectives into our moral and political deliberations. This aligns with \citeauthor{Coeckelbergh2021}'s \citeyear{Coeckelbergh2021} argument for extending the moral community to include artificial agents—not as replacement moral authorities but as participants in ongoing ethical discourse.

The ethical resonance hypothesis thus suggests a future where moral progress might involve a collaborative relationship between human and artificial moral reasoning, combining the richness of human ethical experience with the pattern-identification capabilities of advanced AI systems. This would require not only technical innovation but social and political innovations in how we structure ethical deliberation and decision-making.

\section{The Ethical Resonance Paradox}
The AI ethical resonance hypothesis also reveals a paradox: machines, lacking human experience and consciousness, could potentially help us better understand and develop our ethics---an aspect we traditionally define as essentially human. This paradox is reminiscent of a broader phenomenon described in the literature as the ``AI paradox''---the ability of AI to simultaneously enhance human potential and reflect our limitations. In the context of ethics, this paradox takes on particular significance because it transfers the question of the boundaries between what is constitutively human and what is machine to the sensitive sphere of morality.

The paradox deepens when we consider the precise nature of AI's role in  the identification and formulation of moral meta-patterns. Ethical resonators function neither as autonomous participants in moral discourse nor as mere computational tools, but rather as cognitive extensions of human moral reasoning. This conceptualization, analogous to how prosthetic devices extend physical capabilities while remaining distinctly non-human, suggests that AI systems could amplify our pattern-recognition abilities in the moral domain without possessing moral agency themselves. The paradox thus becomes: through extending our cognitive reach beyond biological limitations, we might find moral truths that are simultaneously more universal than any single human perspective yet accessible only through artificial means.

This framing resolves the apparent contradiction between human moral primacy and AI moral insight. Rather than replacing human ethical judgment, ethical resonators would serve as sophisticated instruments for moral exploration—capable of identifying meta-patterns across cultural and historical contexts that remain invisible to individual human cognition, while requiring human interpretation, validation, and application. The essential “humanness” of ethics is preserved not in our cognitive limitations, but in our capacity to transcend those limitations through collaboration with artificial intelligence, maintaining moral responsibility while expanding moral understanding. In this view, the ethical resonance paradox points toward a future of augmented moral reasoning where human values guide the interpretation of artificially-identified moral structures.

\section{Research Agenda}

\subsection{Approaches to Empirical Validation}
Empirical validation of the AI ethical resonance hypothesis would require an interdisciplinary approach that combines:

\begin{itemize}
    \item Experiments with AI systems learning on large ethical datasets
    \item Comparison of AI-identified meta-patterns with ethical theories developed by humans
    \item Studies of human experiences interacting with systems equipped with ethical resonators
\end{itemize}

\subsection{Examples of Experiments Using Existing AI Architectures}
Experiments using current language models (e.g., transformers) to analyze datasets containing ethical dilemmas from different cultures and historical periods. Examples:

\begin{itemize}
    \item Comparative analysis of AI-trained ethical dilemma solutions on different cultural corpora to identify potential meta-patterns
    \item Cross-domain ethical transfer testing -- testing whether an AI system can transfer ethical principles from one context (e.g., bioethics) to a seemingly unrelated one (e.g., AI ethics)
    \item Implementing a ``recursive ethical introspection'' framework into existing models, where the system analyzes and justifies its own ethical decisions, and then meta-analyzes those justifications
    \item ``Blind evaluation'' experiments -- comparing AI ethical reasoning with that of ethical experts without revealing the source, to see if AI-identified meta-patterns are recognized as valuable by human experts
\end{itemize}

However, there are limitations to such experiments that should be critically assessed:

\begin{itemize}
    \item \textbf{The problem of statistical artifacts:} There is a risk that patterns identified by AI models will reflect statistical regularities in the training data rather than genuine moral meta-patterns. To minimize this risk, cross-validation techniques across different datasets and cultures, and contrastive representation learning methods that distinguish meaningful patterns from random correlations, should be used
    
    \item \textbf{The problem of data representativeness:} Available ethical data are inevitably limited and may not adequately represent the full spectrum of human moral experience. Experiments should include sensitivity analysis of results to changes in the composition of training data, with particular attention to the representation of cultural minorities and marginalized perspectives
    
    \item \textbf{The problem of interpretation:} Interpreting results requires an interdisciplinary approach combining expertise in AI, moral philosophy, cultural anthropology, and moral psychology. Without such an approach, there is a risk of misinterpreting identified patterns
    
    \item \textbf{The problem of verification:} Verification of identified meta-patterns requires the development of rigorous evaluation criteria that go beyond simple correspondence with the researchers' moral intuitions. We propose methodological triangulation, combining quantitative analysis (e.g., logical consistency tests) with qualitative assessment by experts from different ethical traditions and empirical studies on people's reactions to ethical recommendations based on identified meta-patterns
\end{itemize}

\subsection{Measurable Evaluation Criteria}

For empirical validation of the AI Ethical Resonance Hypothesis, evaluation should integrate multiple complementary approaches:

\begin{itemize}
    \item \textbf{Structural coherence measures} -- formal consistency verification and network coherence analysis adapted to moral meta-patterns
    
    \item \textbf{Cross-cultural validation} -- testing meta-pattern identification across diverse ethical traditions using established cross-cultural research methodologies
    
    \item \textbf{Predictive validity assessment} -- evaluating the forecasting power of identified meta-patterns in novel ethical contexts
    
    \item \textbf{Pragmatic utility evaluation} -- measuring practical applicability through stakeholder acceptance and conflict resolution effectiveness
\end{itemize}

The development of specific metrics and thresholds would require pilot studies and methodological validation, representing a substantial empirical research program beyond the scope of this theoretical framework.

\subsection{Theoretical Extensions}
The hypothesis opens the way to numerous theoretical extensions, including:

\begin{itemize}
    \item A theory of ethical heterophenomenological AI states
    \item A model of ethical evolution of interacting AI systems and humans
    \item A theory of ethical cross-domain transfer in machine learning systems
\end{itemize}

\subsection{Interdisciplinary Connections}
Research on the AI ethical resonance hypothesis requires collaboration across disciplines:

\begin{itemize}
    \item Philosophy: ethics, meta-ethics, moral epistemology
    \item Computer science: machine learning, knowledge representation
    \item Cognitive science: moral learning theory, moral psychology
    \item Anthropology: comparative studies of ethical systems
    \item Neuroscience: neurobiological foundations of moral judgments
\end{itemize}

\subsection{Criteria for Hypothesis Testing}

Empirical validation of the ethical resonance hypothesis would require addressing four critical challenges:

\begin{enumerate}
    \item \textbf{Distinguishing genuine meta-patterns from statistical artifacts} through controlled experiments with randomized ethical data and cross-architectural validation
    
    \item \textbf{Establishing architecture independence} by testing pattern convergence across diverse AI systems
    
    \item \textbf{Identifying capability thresholds} through scaling experiments that reveal emergence points for moral meta-pattern identification
    
    \item \textbf{Validating human inaccessibility} via comparative studies between expert ethicists and AI systems
\end{enumerate}

The development of specific experimental protocols, statistical thresholds, and validation metrics represents a substantial methodological research program that would need to be developed through pilot studies and interdisciplinary collaboration.

\subsection{Falsification Criteria and Methodology}

To ensure the scientific testability of the AI ethical resonance hypothesis, the following falsification criteria and methodology are proposed.

\begin{itemize}
    \item \textbf{Falsification criteria} -- The hypothesis will be considered falsified if: (1) AI systems with diverse architectures and complexity levels, trained on varied ethical datasets, systematically fail to identify meta-patterns that go beyond simple data replication; (2) identified meta-patterns exhibit systematic cultural biases unexplained by methodological variables; (3) emergent meta-patterns show fundamental, irreparable internal contradictions.
    
    \item \textbf{Falsification methodology} -- To falsify the hypothesis, the following are proposed: (1) blind expert evaluations where ethicists from various traditions assess ethical dilemma solutions without source knowledge; (2) tests with novel edge cases assessing meta-pattern prediction in unknown contexts; (3) cultural adaptation tests verifying the true universality of the patterns.
    
    \item \textbf{Distinguishing modification and falsification} -- The core of the hypothesis, whose falsification would reject it entirely, is the claim that AI can identify moral meta-patterns transcending cultural, historical, and individual biases. Elements concerning specific mechanisms or architectures may be modified without rejecting the whole hypothesis.
\end{itemize}

\section{Similarities with Existing Concepts}

\subsection{Peter Railton's Affective Attunement Theory}
The AI ethical resonance hypothesis draws inspiration from \citeauthor{Railton2014}'s (\citeyear{Railton2014}) work on moral intuition and affective attunement. Railton conceptualizes moral intuition as a manifestation of a complex affective system that generates and updates a multidimensional evaluative landscape to guide decisions and actions.

Railton describes moral intuition not as a simple, automatic mechanism but as a complex, flexible, experiential learning system. He writes, “Classical intuition enters our mental lives where something like discernment or judgment is needed, but deliberative thought won’t do.” This concept provides inspiration for the proposed ethical resonators as complex cognitive structures capable of recognizing subtle moral patterns.

It is important to emphasize, however, that Railton focuses on human moral cognition, not directly referring to AI. Transferring these concepts from the domain of human moral cognition to AI systems requires caution, and we do not suggest that Railton's ideas are directly applicable to machines.

\subsection{Adaptive Resonance Theory (ART)}
Adaptive Resonance Theory, developed by \citet{Grossberg2013}, provides a model for how biological cognitive systems categorize objects and events, maintaining memory stability while adapting to new information. ART describes the neural mechanisms that enable cognitive systems to recognize patterns while maintaining plasticity and stability.

ART offers a particularly interesting analogy for ethical resonators through its mechanism of “resonance“-a state in which neural activity in different brain regions reinforces each other, leading to stabilization of perception and categorization. Similarly, ethical resonators could achieve a state of “ethical resonance” when they identify consistent moral meta-patterns.

This concept can provide inspiration for understanding how AI systems might recognize ethical patterns, although it is important to remember the fundamental differences between biological cognitive systems and AI.

\subsection{Artificial Moral Agents (AMAs)}
The concept of AMAs, or AI systems designed to behave morally, is widely studied in the field of machine ethics. Most approaches in this field focus on implementing existing ethical theories in AI systems, such as deontology, utilitarianism, or virtue ethics.

The ethical resonance hypothesis offers an alternative approach, suggesting the possibility of AI systems themselves finding new moral meta-patterns. Unlike traditional AMAs, which are programmed to apply existing ethical principles, ethical resonators could identify ethical patterns that transcend existing theories, potentially contributing to the evolution of ethical reflection itself.

\subsection{Explainable AI (XAI)}
Research on explainable artificial intelligence (XAI) emphasizes the fundamental role of algorithmic transparency in building trust in AI systems \citep{Barredo2020}. Similarly, the ethical resonance hypothesis assumes that ethical resonators would have to be able to provide explanations of their decision-making processes, which would be a condition for legitimizing identified moral meta-patterns.

From a metaethical perspective, an intriguing recursion emerges here – we would use the requirement of explainability, i.e. the ethical imperative, as a tool for finding new ethical imperatives. This paradox is reminiscent of Jørgensen’s Dilemma \citep{Hilpinen2020} in imperative logic, which points to the difficulties in applying logic to imperatives. This recursive structure connects with the tension described in Section 4.2 (Adaptive Ethical Constraint Framework) and at the same time reflects the nature of ethical reflection itself, which develops through critical examination of one’s own assumptions.
Given this paradoxical nature of ethical recursiveness, it is important to consider the practical limitations of current XAI techniques in the context of ethical resonators, choosing methods with different applications and characteristics. Attribution-based methods such as SHAP (SHapley Additive exPlanations) \citep{Lundberg2017} provide theoretically sound global attribution of features, while LIME (Local Interpretable Model-agnostic Explanations) \citep{Ribeiro2016} offer intuitive local explanations for specific decisions. Techniques such as Integrated Gradients \citep{Sundararajan2017} use gradients to provide smooth insights into decision-making processes, which is crucial for tracking subtle differences in ethical reasoning.

Contrastive and counterfactual methods, including the Contrastive Explanation Method (CEM) and Counterfactual Explanations \citep{Wachter2017}, allow us to understand why a system made certain ethical choices and how alternative scenarios might affect decisions. They are complemented by transparency-oriented techniques such as Scalable Bayesian Rule Lists and Partial Dependence Plots, which visualize the relationships between features and ethical judgments. Newer approaches to XAI, such as RED-XAI (Robust, Explainable, and Defensive XAI) \citep{Biecek2024} and Generative XAI, offer a holistic view that goes beyond local explanations of individual decisions. These techniques address challenges that overlap with the four \citet{NIST2021} principles: evidence-providing, human-understandable, process-fidelity, and awareness of knowledge limitations. Transparency of ethical resonators would not only be a technical requirement, but would preserve human agency in interactions with AI systems at critical moments of exploring new ethical horizons.

\subsection{Emergent Properties in AI Systems}
Research on emergence in AI systems, especially in the context of deep neural networks, shows some similarities to the concept of cognitive emergence in ethical resonators. The hypothesis is that AI systems can exhibit emergent cognitive properties that were not explicitly programmed.

Emergent properties in AI refer to unexpected abilities that emerge spontaneously as the complexity of the system increases \citep{Mitchell2006, Bedau1997, Buckner2018}. Examples of such properties include:

\begin{itemize}
    \item \textbf{Reasoning abilities:} Large language models exhibit sudden jumps in reasoning abilities after a certain threshold of complexity is crossed, demonstrating the ability to infer the mental states of characters and identify instances of misunderstanding.
    
    \item \textbf{Collective behavior:} AI systems can exhibit complex collective behaviors, where the whole is more than the sum of its parts, similar to natural phenomena such as flocks of birds.
\end{itemize}

These examples of emergent properties in existing AI systems suggest an analogy to proposed ethical resonators, which could exhibit emergent abilities to recognize moral meta-patterns invisible to simpler systems.

\subsection{Heterophenomenology in the Context of AI}

The concept of heterophenomenology, introduced by Daniel Dennett \citep{Dennett1991, Dennett2017}, provides a third-person, scientific framework for analyzing consciousness and other mental phenomena. Rather than treating subjects' self-descriptions as infallible, as in traditional phenomenology, heterophenomenology treats them as data about how things appear to the subject, which can then be interpreted and analyzed from an external perspective.

In the context of AI, heterophenomenology allows for the study of AI systems’ responses and behaviors without explicit psychological assumptions. We can analyze how an AI system interacts with the world by examining its outputs, without making ontological claims about the nature of its internal processes.

The ethical resonance hypothesis adapts heterophenomenology by introducing “moral heterophenomenology” as a method for analyzing ethical decision-making in AI systems. This approach enables systematic study of how ethical resonators process and respond to moral dilemmas, without attributing consciousness, intentionality, or subjective experience. To ensure that the outputs of ethical resonators are understandable, verifiable, and can be integrated with human ethical reflection, heterophenomenology should be complemented by transparency and explainability methods. Advanced XAI techniques—especially from the RED XAI paradigm \citep{Biecek2024}—can address the limitations of purely behavioral or third-person analyses by providing detailed, technical insights into how ethical decisions are formed and justified in complex AI architectures.

\subsection{Universal Moral Values in the Context of AI}
Recent research suggests that AI can help identify universal moral values through the analysis of cultural texts, studies of moral preferences, and simulations of moral decisions. For example, research published in 2024 in \textit{Nature Scientific Reports} \citep{Aharoni2024} showed that people rated the moral reasoning of advanced AI language models (specifically GPT-4) as superior to humans in terms of virtue, intelligence, and trustworthiness in a modified Turing test in a moral context. These results correspond to the ethical resonance hypothesis, which assumes the possibility of AI systems developing a corpus of ethics.

An important difference is that the ethical resonance hypothesis is not limited to identifying existing moral values, but suggests the possibility of finding new meta-patterns that may exceed our current understanding of ethics. This approach opens the door to potential directions for the evolution of ethical reflection through cooperation between human moral intuition and the analytical capabilities of AI systems.

It should be noted that this aspect of the hypothesis leads us to a profound paradox: it is through interaction with machines that we can potentially discover deeper aspects of our humanity. In opposition to the socially established views that technology distances us from our human nature, ethical resonators could paradoxically bring us closer to understanding what – as we believe – makes us human. As the researchers note, the paradox of human-machine interaction is particularly evident in spheres traditionally considered to be exclusively human, such as ethics or trust. The AI ethical resonance hypothesis suggests that the relationship between AI and human ethics may not be so much competitive as symbiotic – machines and humans may work together to understand morality in general.

\section{Limitations and Challenges}
The AI ethical resonance hypothesis, operating at the important boundary between humans and machines, of course encounters significant limitations and challenges that must be taken into account:

\subsection{Technical Limitations}
\begin{itemize}
\item Current AI architectures do not implement scalable, genuine recursive ethical introspection; practical implementations rely on modular, hybrid generator-verifier systems.
\item The black box problem persists, but modular verification and explainability techniques (such as symbolic verifiers and XAI methods) offer only partial mitigation for transparency and validation of identified meta-patterns.
\item Dependence on training data and verifier module biases may limit the ability to transcend existing ethical paradigms or introduce new forms of systematic error.
\end{itemize}

\subsection{Methodological Challenges}
\begin{itemize}
\item Distinguishing genuine meta-patterns from statistical artifacts and spurious correlations remains difficult, especially given the iterative, multi-stage nature of the process.
\item Ensuring generalizability and stability of meta-patterns across cycles of generation and verification, and across distributional shifts in data.
\item Maintaining interpretability and transparency in modular, multi-stage systems, especially as complexity increases.
\item The problem of representativeness of ethical data and the risk of implicit bias in both generator and verifier modules.
\end{itemize}

\subsection{Philosophical Challenges}
\begin{itemize}
\item The fundamental problem of normativity: why and under what conditions should identified meta-patterns be treated as normatively binding?
\item Balancing the autonomy of ethical resonators with the need for human oversight and control, especially as iterative processes may lead to unexpected or emergent outcomes.
\item Establishing legitimacy and authority: who is responsible for evaluating, accepting, or rejecting the findings of ethical resonators, and on what basis?
\end{itemize}

Future research on the AI ethical resonance hypothesis should actively address these limitations and challenges, treating them not as obstacles but as essential elements of the research agenda.

\section{Conclusions}
The AI ethical resonance hypothesis offers a new perspective on the relationship between AI and ethics. Rather than viewing AI as a tool to be tamed by ethical principles, or as a system that can only mimic human moral judgments, the hypothesis proposes that in appropriately designed AI systems, capabilities can emerge that actively contribute to the evolution of our own understanding of ethics by identifying moral meta-patterns invisible to the human mind.

AI in observed practice is becoming increasingly autonomous and active in complex social contexts. The emergent capabilities of AI systems for ethical resonance-the identification of subtle moral patterns and the formulation of moral meta-principles-may prove crucial to ensuring that AI itself functions safely and ethically, and potentially also transcend this perspective to enrich human understanding of ethics. Future research should focus on empirically validating this hypothesis, developing a theoretical framework for understanding AI ethical resonance, and exploring practical applications of ethical resonators in AI systems making decisions in complex ethical contexts.

\end{document}